\documentstyle[12pt]{article}
\begin{document}
lgonzalz@vxcern.cern.ch
~~~~~~~~~~~~~~~~~~~~~~~~~~~~~~~~~~~~~~~~~~~~~~~~~~~~~~~~~~~~~~~~~~~~~~~~April
1997
\vskip 2.4cm
\centerline {\bf VACUUM STRUCTURE, LORENTZ SYMMETRY}
\vskip 3mm
\centerline {\bf AND SUPERLUMINAL PARTICLES (I)}
\vskip 2cm
\centerline {\bf L. GONZALEZ-MESTRES}
\vskip 5mm
\centerline {Laboratoire de Physique Corpusculaire, Coll\`ege de France}
\centerline {11 pl. Marcellin-Berthelot, 75231 Paris Cedex 05 , France}
\vskip 3mm
\centerline {and}
\vskip 3mm
\centerline {Laboratoire d'Annecy-le-Vieux de Physique des Particules}
\centerline {B.P. 110 , 74941 Annecy-le-Vieux Cedex, France}
\vskip 2.9cm
{\bf Abstract}
\vskip 4mm
If textbook Lorentz invariance is actually
a property of the equations describing a sector 
of the excitations of vacuum above some critical distance scale,
several sectors of matter with different
critical speeds in vacuum can coexist and an absolute rest frame (the vacuum
rest frame)
may exist without contradicting the apparent Lorentz invariance felt by
"ordinary" particles (particles with critical speed in vacuum equal to $c$ ,
the speed of light). Sectorial Lorentz invariance, reflected by the fact that
all particles of a given dynamical sector have the same critical speed in 
vacuum, will then be an expression of a fundamental sectorial symmetry
(e.g. preonic grand unification or extended supersymmetry) protecting a 
parameter of the equations of motion. Furthermore, the sectorial Lorentz
symmetry may be only a low-energy limit, in the same way as the relation
$\omega $ (frequency) = $c_s$ (speed of sound) $k$ (wave vector) holds for
low-energy phonons in a crystal.
We study the consequences of such a scenario, using an ansatz inspired by 
the Bravais lattice as a model for some vacuum properties. It then turns out
that: a) the Greisen-Zatsepin-Kuzmin cutoff on high-energy cosmic protons and 
nuclei does
no longer apply; b) high-momentum unstable particles have longer lifetimes
than expected with exact Lorentz invariance, and may even become stable
at the highest observed cosmic ray energies or slightly above. 
Some cosmological 
implications of superluminal particles are also discussed.
\vskip 20cm
\vskip 1cm
{\bf 1 . VACUUM EXCITATIONS AND "ELEMENTARY" PARTICLES}
\vskip 5mm
Sparnay and Lamoreaux [1] have experimentally confirmed the Casimir effect
[2 - 4] based on the quantum-field-theoretical interpretation of 
elementary particles,
which states that quantum fields are harmonic oscillators in vacuum and have 
therefore a nonzero ground energy. Thus, the so-called "elementary particles"
are actually quantum oscillators of vacuum degrees of freedom and 
conventional quantum field theory is based on the harmonic approximation to
these oscillators. Quantum mechanics seems  
to arise at a deeper level than the
description of the so-called "free particles".
This raises a fundamental question: can
matter be understood just from a phenomenological study of the excitations of
vacuum? Obvioulsy, the matter forming a ionic crystal cannot be described just
in terms of the crystal phonon spectrum. On the other hand, our present
knowledge of vacuum excitations (quarks, leptons, gluons,
electroweak gauge bosons,
string models...)
is likely to contain, if correctly interpreted, important information on 
vacuum dynamics itself. A most remarkable fact is that all the above-mentioned
excitations of vacuum seem to have the same critical speed in this medium.
The dynamical origin of
such a symmetry is far from trivial if we adopt the philosophy that the 
apparent structure of space and time, as seen by matter, is actually 
determined by the properties of matter itself (e.g. [5 - 11]). This would
not really be an unorthodox approach, as 
standard (inflationary) 
cosmology generates space from the expansion (creation) of matter.
\vskip 5mm
{\bf 1a. Vacuum and particles: an analogy with the Bravais lattice}
\vskip 5mm
To understand, in the absence of absolute prescriptions from an intrinsic
space-time geometry, the meaning of this apparent universality of the critical
speed in vacuum, we can attempt a simple analogy with solid state physics.
Assume a classical system similar to the 
monoatomic one-dimensional Bravais lattice [12]  
with a very large number of coupled oscillators regularly spaced 
by $a$ , $\pi $ times
the inverse of the critical wave vector $k_0$
($k_0$ and $-k_0$ leading to the same wave function).
On each site $n$ , a complex 
parameter $\phi (n)$ satisfies an equation implying nearest-neighbour
coupling, i.e. :
\equation
d^2/dt^2~~[\phi ~(n)]~~~=~~~K~[2~\phi ~(n)~-~\phi ~(n-1)~-~\phi ~(n+1)]~~-~~
\omega _{rest}^2~\phi
\endequation  
\noindent
which admits, for wave vector $k$ 
($-k_0~\leq ~k~\leq ~k_0$), solutions of the type:
\equation
\phi _k(n~,~t)~~=~~\phi _k~(0)~~exp~[i~(k~n~a~-~\omega ~t)]
\endequation
\noindent
with 
\equation
\omega ^2~(k)~~=~~2~K~[1~-~cos~(k~a)]~+~\omega _{rest}^2~~=
~~4~K~sin^2~(ka/2)~+~\omega _{rest}^2
\endequation
which, at low $k$ and taking positive energy solutions, can be written as:
\equation
\omega ~(k)~~\simeq ~~[K~(a~k)^2~+~\omega _{rest}^2]^{1/2}
\endequation
\par
\noindent
as in standard special relativity. We expect $\omega _{rest}~=~0$ as long as 
global fluctuations have no energy. 
Each plane wave $\phi _k(n~,~t)$ can be viewed as a complex 
harmonic oscillator.
The speed $d\omega /dk$ is basically 
determined by the spring constant $K$ and the
critical wavelength $k_0$ = $\pi~a^{-1}$ . With the above approximations, 
and interpreting the lattice as the physical vacuum, a 
classical massive field is obtained which can be quantized to give a 
charged massive 
particle. 
By this simple procedure, a particle has been generated whose
critical speed in vacuum is explicitly related to dynamical parameters
of this vacuum
($K$~,~$a$). If, as it seems to occur in our world, many particles have the
same critical speed in vacuum, they must correspond to oscillators with the
same values of these dynamical parameters. The observed symmetries 
must correspond to symmetries of the inner vacuum dynamics
(preonic symmetry, supersymmetry...) and may survive
beyond the scales where the particles cease to exist. Instead of a complex
parameter, we can take in the previous example
a unitary operator $u~(n~,~n-1)$ 
associated to links between 
sites. If $u~(n~,~n-1)$ belongs to the unitary group 
$U~(N)$ , $\lambda $ is a hermitic generator of $U~(N)$ and
$\alpha $ a real parameter, the plane wave:
\equation
u _k~(n~,~n-1~;~t)~~=~~exp~[i~(k~n~a~-~\omega ~t)]~exp~(i~\alpha ~\lambda)
\endequation
satisfies the equation:
\equation
d^2/dt^2~~[u~ (n~,~n-1)]~~~=~~~K~[2~u~ (n~,~n-1)~-
~u~ (n-1~,~n-2)~-~u~ (n+1~,~n)]
\endequation
for:
\equation
\omega ^2~(k)~~=~~2~K~[1~-~cos~(k~a)]~~=~~4~K~sin^2~(ka/2)
\endequation
\par
The hermitic vector
field, which satisfies (6) and (7) similar to $u$ , is actually:
\equation
A~(n~,~n-1)~~=~~(2i)^{-1}~[u~(n~,~n-1)~-~u~(n-1~,~n)]
\endequation
where $~u~(n-1~,~n)~~=~~u^{\dagger }~(n~,~n-1)$ . $A$ is hermitic and contains
both positive and negative frequencies:
\equation
2~A_k~(n~,~n-1)~~=~~a^{\dagger }_k~exp~[-i~(k~n~a~-~\omega ~t)]~+
~a_k~exp~[i~(k~n~a~-~\omega ~t)]
\endequation
where $a_k~=~-i~exp~(i~\alpha ~\lambda)$ . In this way, nine 
massless bosons can be generated, as $\lambda $ varies, 
possibly equivalent to a set of
$U(N)$ gauge bosons.  
Matter fields will exhibit (up to a mass) 
the same relation between energy and momentum as gauge fields
if they 
oscillate with the same parameters $K$ and $a$
as the internal symmetry links,
which can be a natural assumption if in both cases we are dealing with
nearest-neighbour interactions of the same family of degrees of freedom. 
In this scenario, the violation of Lorentz symmetry
introduced at high energy by equations (3) and (7) 
will not be accompanied by a
breaking of the universality of the relation between energy and momentum
in the zero-mass limit. Although masses and other related
phenomena will introduce
low-energy corrections to this universality, the high-momentum dynamics can 
keep it unbroken. An ansatz for a general formula describing 
kinematics in the vacuum rest frame
can be as before: 
\equation
\omega ^2~(k)~~=~~2~K~[1~-~cos~(k~a)]~+~(2\pi)^2~h^{-2}~E_{rest}^2
\endequation
where $h$ is the Planck constant and $E_{rest}$ the rest energy of the particle.
The "speed of light", as measured at low energy
in the limit $k~\rightarrow ~0$ , is $c~=~K^{1/2}~a$ . The inertial  
mass of the particle in the same limit is, as usual, $m~=~E_{rest}~c^{-2}$ .   
\vskip 4mm
In such an analogy, no basic principle would prevent the vacuum from having
other sets of
degrees of freedom oscillating with different values of $a$ and $K$ .
These degrees of freedom may generate superluminal sectors of matter, as
considered in [5~-~11] . If $K_i$ and $a_i$ are the values of $K$ and $a$
for the $i$-th superluminal sector, we may try for these superluminal 
particles in the vacuum rest frame the ansatz:
\equation
\omega ^2~(k)~~=~~2~K_i~[1~-~cos~(k~a_i)]~+~(2\pi)^2~h^{-2}~E_{rest}^2
\endequation
where, from [5~-~10] , the mass of the particle is related to its rest
energy by the relation $E_{rest}~=~m~c_i^2$ ,
and $c_i~=~K_i^{1/2}~a_i~=~\pi ~K_i~k_i^{-1}$ , 
where $k_i$ is the critical
wavelength. 
Interaction between different dynamical
sectors would then lead to energy and momentum transfer between different kinds
of excitation modes. As stressed in [5~-~11] , the superluminal particles 
we consider would
not be tachyons and would have quite different physical properties from those
predicted for tachyonic objects [13] . 
Contrary to tachyons, the new superluminal particles necessarily
violate the relativity principle [14] and would produce "Cherenkov" radiation
(spontaneous emission of particles with lower critical speed)
in vacuum (see [5~-~11] , and also [15]). 
They can play an 
important role in high-energy phenomena, as well as in cosmology,
and yield detectable signatures 
at accelerators (e.g. LHC) or in
cosmic ray experiments (e.g. AMANDA [16]).
\vskip 4mm
Obviously, kinematics from (10) and (11)
is just a rough example parameterizing possible 
trends of scenarios where Lorentz invariance is only a low-energy limit 
and "ordinary" particles (those with critical speed in vacuum equal to $c$ , 
the speed of light) cease to exist at distance scales 
below $a$ . However, it may allow for useful discussions
of Lorentz symmetry violation and of other related phenomena. Consequences 
in cosmology can be important, especially for the Big Bang scenario [17~,~18] . 
Quantum field theory [19~-~21] should take these penomena into 
account, especially when discussing cutoffs and renormalization, 
but its validity is not in principle altered by their existence. Contrary to
tachyons [22~,~23] , the proposed superluminal particles would not violate
causality [10] . From formula (10) , the 
velocity of an "ordinary" particle would
be ($E$ = energy, $p$ = momentum):
\equation
v~~=~~dE/dp~~=~~d\omega /dk~~=~~K^{1/2}~c~\omega ^{-1}~sin~(k~a)
\endequation 
which tends to zero 
as $k$ approaches $\pm \pi a^{-1}$ . 
Thus, a particle with the highest permitted 
energy and wave vector would be at rest (zero speed) with respect to the
vacuum rest frame, even if it has a high momentum.
In this limit,
the frequency tends to 
$\omega ^{max}~~=~~[4~K~+~(2\pi)^2~h^{-2}~E_{rest}^2]^{1/2}$ .
Similar expressions can be obtained
for superluminal particles.
In the early Universe, the characteristic phase transition temperature
sclaes [2~,~9] would be: 
\equation
T_0~~\approx ~~k_B^{-1}~h~K^{1/2}
\endequation
for the ordinary sector, $k_B$ being the Boltzmann constant, and:
\equation
T_i~~\approx ~~k_B^{-1}~h~K_i^{1/2}
\endequation
for each superluminal sector. The cosmological phase transition 
temperatures depend, for each sector, on the sectorial spring constant and not
on the critical wavelength scale. An interesting scenario would be, for all 
$i$ : $a_i~=~a$ ; $K_i~\gg ~K$ . Then, the "Planck scale" may still make sense
as a distance scale, but the "Planck temperature" (equivalent to a 
Debye temperature)
would be different for different sectors. Above $T_0$ ,
only superluminal particles would in principle exist
and "Cherenkov" radiation in vacuum may have been inhibited by the very short
time scale. 
In this case, the Universe would 
have quickly cooled down just after having reached each critical
temperature ($T_0$ and the $T_i$'s). 
Another
possible scenario would be: $a_i~\gg~a$ ; $c_i~\gg ~c$ which requires very
large values of $K_i$ but is not prevented by any basic principle. 
In such case, a
question arises: since superluminal particles remain "elementary"
at scales where the 
internal structure of ordinary particles 
(quarks, leptons, gauge bosons...) shows up, are the former
part of the constituents of the vacuum structure generating the latter?
If the answer is positive
for one of the superluminal sectors, 
transparency of vacuum with respect to superluminal
particles of this sector
would not be a trivial problem. Superluminal particles 
raise several fundamental questions concerning their origin and properties
as possible components of vacuum dynamics. 
\vskip 5mm
{\bf 1b. Further analogies}
\vskip 5mm
The analogy with solid state physics raises in itself two interesting questions:
\vskip 3mm
- Do $umklapp$ processes [12] take place at very high energy? In such 
processes, the total momentum would be conserved only up to an integer
multiple of $h~a^{-1}~{\vec {\bf u}}$ 
for ordinary particles, and of $h~a_i^{-1}~{\vec {\bf u}}$ for superluminal 
particles, where ${\vec {\bf u}}$ is a
unitary vector. This (or a more involved scenario) cannot be 
excluded, as there is no guarantee that translation invariance remains
valid  below the $a$ and $a_i$ distance scales.  
Dynamics would then depend crucially on the topology of momentum space.
This topological space can be the product of three circles like in a crystal
lattice. But it can also be, for instance, a sphere with antipodes identified
on its surface (i.e. $k_0~{\vec {\bf u}}~=~-~k_0~{\vec {\bf u}}$ for any 
unitary vector ${\vec {\bf u}}$). 
Then, rotation invariance would be preserved and expression (10) can naturally
apply to three space dimensions.
We can also consider a 
momentum topology where all points on the surface 
$\mid {\vec {\bf u}}\mid ~=~k_0$
of the momentum sphere
would be identified to a single point (making the momentum space topologically
similar
to the $SU(2)$ group). 
$Umklapp$ processes would
modify the thermal conductivity of the very early Universe, and make more
difficult heat exchanges between different regions of 
space at high temperature.
\vskip 3mm
- Is the "acoustic branch" [12] the only dynamical branch of ordinary or 
superluminal particles? The question is not merely academic, as the contrary
may imply the existence of "optical"
particles with: a) negative inertial mass at zero momentum, but
positive rest energy; b) minimum energy at maximum momentum.
Such particles 
would be very heavy and, at high energy, 
undergo a repulsive acceleration in the presence
of a static, attractive, gravitational feld.
They may have crucially influenced the expansion of the Universe.
"Optical" particles can be generated replacing in 
the previous analogy the monoatomic lattice by a diatomic one. 
We would then be
led in the vacuum rest frame, in a similar way to [12] in one space dimension 
and assuming rotation invariance when generalizing the ansatz
to the three-dimensional case, 
to the frequency spectrum:
\equation
\omega ^2~~=~~K_0~\pm ~K_1~(k)
\endequation
with:
\equation
K_0~~=~~K~+~G~+~\lambda _1 ~+~\lambda _2
\endequation
\equation
K_1~~=~~[K^2~+~G^2~+~2~K~G~cos~(ka)~+
~(\lambda _1 ~-~\lambda _2)^2]^{1/2}
\endequation
\noindent
where $2\lambda _1$ , $2\lambda _2$ are monoatomic elastic constants similar
to $\omega _{rest}^2$ in the previous example and $K$ , $G$ govern 
nearest-ion interactions [12] . All these constants are positive in realistic 
examples based on harmonic oscillators.
The solution with the $-$ sign ("acoustic" particles)
can be dealt with in a similar way
to the monoatomic case, and leads to a massless particle if
$\lambda _1 ~=~\lambda _2~=~0$ . Writing $\delta ~=~
\lambda _1 ~-~\lambda _2$ , the solution with a $+$ sign 
(the "optical branch") has rest (zero momentum) frequency:
\equation
E_{rest}~(optical)~~=~~(2\pi )^{-1}~h~ 
[K_0~+ ~(K^2~+~G^2~+~2~K~G~+~\delta ^2)^{1/2}]^{1/2}
\endequation
\noindent
and negative values of $d^2\omega /dk^2$ . In the limit where $k$ approaches
$\pm \pi ~a^{-1}$ , the energy approaches the positive value:
\equation
E_{min}~(optical)~~=~~
(2\pi )^{-1}~h~[K_0~+~
(K^2~+~G^2~-~2~K~G~+~~\delta ^2)^{1/2}]^{1/2}
\endequation
and the particle would also be at rest (zero speed) in this limit,
where it has its minimum possible energy and its maximum permitted momentum.
Since this is the less energetic state, it is likely that "optical"
particles tend to be in such a state in the present epoch.
There, for ${\vec {\bf k}}~=~k_0~{\vec {\bf u}}$ where ${\vec {\bf u}}$
is an arbitrary unitary vector,
they would have a large, positive inertial mass in the direction of their 
momentum 
and infinite inertial mass in the two other directions.
More precisely, we can write for an "optical" particle around maximum 
wavelength ${\vec {\bf k}}~=~k_0~{\vec {\bf u}}$ submitted to a static
external force the hamiltonian:
\equation
H~~=~~(2\pi)^{-1}~h~[K_0~+ ~K_1~(k)]^{1/2}~+~V~({\vec {\bf r}})
\endequation
where ${\vec {\bf r}}$ is the position vector
and, with the relation ${\vec {\bf~k}}
~=~2\pi~h^{-1}~{\vec {\bf~p}}$ where ${\vec {\bf~p}}$
is the momentum, the classical Hamilton equations:
\equation
{\vec {\bf v}}~~=~~2\pi ~h^{-1}~{\vec {\bf {\nabla _k}}}~H~~=~~
-~(a~k^{-1}/2)~[K_0~+ ~K_1~(k)]^{-1/2}~sin~(ka)~{\vec {\bf k}}~dK_1/d[cos(ka)] 
\endequation
\equation
d~{\vec {\bf k}}/dt~~=~~-~2\pi~h^{-1}~{\vec {\bf {\nabla }}}~V~({\vec {\bf r}})
\endequation
where ${\vec {\bf {\nabla _k}}}$ stands for gradient in wave vector space. 
For a unitary vector ${\vec {\bf u}}$ :
\equation
d~{\vec {\bf v}}/dt\mid _{{\vec {\bf k}}~=~k_0{\vec {\bf u}}}~~=~~
-~[2~E_{min}~(optical)]^{-1}~K~G~[K~(k_0)]^{-1}~a^2~
[{\vec {\bf u}}.{\vec {\bf {\nabla }}}~V~({\vec {\bf r}})]~{\vec {\bf u}}
\endequation
\par
Therefore, an "optical" particle at minimum energy
can be accelerated only by a force parallel 
to its momentum. 
A condensate of "optical" particles in vacuum may then spontaneously break
rotation invariance, as "optical" particle-antiparticle pairs would be at
rest with maximum momenta pointing, locally, in arbitrary directions.
The observed isotropy of 
cosmic microwave background radiation as well as our daily 
experience and the long range of 
gravitational forces seem to indicate that such an effect, 
as felt by "acoustic" particles, must be very small when observed 
at large distance scales. At shorter distance scales (e.g. those reached by 
accelerator experiments), it may be worth to perform precision tests of
rotation invariance. 
The threshold for "optical" particle production would be given by the
energy at maximum
wave vector (we expect masses $\approx 10^{19}~GeV~c^{-2}$ for $a~\approx
~10^{-33}~cm$),
and not by the $k~=~0$ rest energy as for "acoustic" particles.
Particles of the "acoustic branch" would be faster than those of the 
"optical branch", but their maximum energy would be lower than the minimum
energy of "optical" particles. "Optical" particles, or instead more complicated 
objects, may indeed appear at very high energy as a consequence of nontrivial
vacuum structure. As they seem to be in "one-to-one" 
correspondence with "acoustic" particles, 
we consider them as being part of the same
dynamical sector.
The question arises whether such "optical" objects,
whatever they are, can play a role in 
the renormalization of quantum field theories. In any case, it seems 
difficult to imagine how "bare" particles could
reasonably be defined without taking
into account physics at the natural cut-off scales
where "point-like" interactions do no longer make sense.
\vskip 4mm
In principle, there is no basic reason for "optical" particles to 
couple to the "acoustic" graviton in the same way as "acoustic" particles.
For instance, the low-momentum Lorentz symmetry 
becomes euclidean and with a different value of the
critical speed. "Optical" particles correspond to fundamentally different
vacuum excitations and would not necessarily decay into "acoustic" ones.
If such decays do not occur, some "optical" particles 
may be stable. Since they are in one-to-one correspondence with
"acoustic" particles, "optical" particles can possibly couple to "acoustic"
internal-symmetry gauge bosons. If ordinary particles
are excitations of vacuum degrees of freedom associated to a condensate of
particles of the $i$-th superluminal sector and involving a long range 
superluminal force, ordinary "optical" particles 
at small wave vector may mix with the superluminal field, in the
same way as optical phonons
mix with the electromagnetic field in a ionic crystal [12]. This would 
give rise to superluminal "polaritons" propagating in vacuum.
\vskip 4mm
If "optical" particles have significant
couplings to "acoustic" 
gravitation and  
internal-symmetry gauge bosons, several interesting phenomena
can be expected. 
Interactions between "optical" and "acoustic" particles would present rather 
unconventional features. 
The low-energy hamiltonian for a system formed by an "acoustic" particle of
mass $m_a$ and an "optical" particle of 
effective inertial mass $M_O$ will be in the vacuum rest frame:
\equation
H~~\simeq ~~p_a^2~(2~m_a)^{-1}~+~(2M_O)^{-1}~(2\pi~a)^{-2}~h^2~sin^2(ka)~
+~V~({\vec {\bf r}})
\endequation
\noindent
where ${\vec {\bf p}}_a$ is the momentum of the 
"acoustic" particle, $m_a$ its mass,
$M_O$ the effective inertial mass of the
"optical" particle, 
${\vec {\bf k}}$ its wave vector,
${\vec {\bf r}}~=~ 
{\vec {\bf r}}_a~-~{\vec {\bf r}}_O$ , ${\vec {\bf r}}_a$
and ${\vec {\bf r}}_O$ the position vectors of the "acoustic" and the "optical" 
particle, and $V~({\vec {\bf r}})$ the potential energy.
If the "optical" particle is close to $k~=~k_0$, we can write:
\equation
sin^2~(k~a)~~=~~sin^2~(k~a~-~k_0~a)~~\approx ~~(k~a~-~k_0~a)^2
\endequation
\equation
H~~\simeq ~~p_a^2~(2~m_a)^{-1}~+~(2M_O)^{-1}~(p_O~-~h~a^{-1}/2)^2~
+~V~({\vec {\bf r}})
\endequation
where ${\vec {\bf p}}_O$ is the momentum of the "optical" particle,
and Hamilton equations lead to:
\equation
dp_O/dt~~=~~-~dp_a/dt~~=~~{\vec {\bf \nabla }}~~V~({\vec {\bf r}})
\endequation
\equation
{\vec {\bf v}}_a~~=~~m_a^{-1}~{\vec {\bf p}}_a
\endequation
\equation
{\vec {\bf v}}_O~~=~~M_O^{-1}~(p_O~-~h~a^{-1}/2)~p_O^{-1}~{\vec {\bf p}}_O
\endequation
where ${\vec {\bf v}}_a$ and ${\vec {\bf p}}_O$ stand for the velocity of
the "acoustic" and "optical" particle. The last equation can be turned into:
\equation
{\vec {\bf p}}_O~~=~~{\vec {\bf v}}_O~v_O^{-1}~(M_O~v_O)~+~h~a^{-1}/2)
\endequation
so that the conserved total momentum ${\vec {\bf P}}$ is given by:
\equation
{\vec {\bf P}}~~\simeq ~~m_a~{\vec {\bf v}}_a~+~M_O~{\vec {\bf v}}~+~
h~a^{-1}~v_O^{-1}~{\vec {\bf v}}_O/2
\endequation 
\par 
It directly follows from this expression that, if the momentum transfer
amounts only to a rotation of ${\vec {\bf p}}_O$ , it will be mainly spent in
the rotation of the zero-speed component of the momentum of the "optical"
particle. Only a fraction 
$\simeq ~2~M_O~v_O~h^{-1}~a$ would in such case lead to 
a modification  the effective momentum $M_O~v_O$ of the "optical" particle.
In the limit where the "optical" particle is initially
at rest, there will be no acceleration for a transverse infinitesimal 
momentum transfer. Thus, naive conservation laws will not apply to
the "effective" total momentum
$m_a~{\vec {\bf v}}_a~+~M_O~{\vec {\bf v}}_O$ .
\vskip 4mm
"Optical" particles may be unconventional dark matter candidates playing an
important role in the formation of structure in the early Universe.
As for "acoustic" particles, annihilation may have been prevented by 
matter-antimatter asymmetry. 
If such particles exist, are stable and have not annihilated, they will be 
present in the 
matter-dominated Universe.
If their density is of the order of
standard cosmological matter densities, and if 
they are practically
at rest with respect to the vacuum rest frame (assumed to be close
to that defined by cosmic microwave background), they will move at a speed
$\approx 2.10^{-3}~c$ with respect to the Local Group. We can then expect
on earth fluxes $\approx 10^{-6}~m^{-2}~year^{-1}$ which would not be
unaccessible to future cosmic ray detectors.
An interesting question is whether they can be found in "acoustic" matter.
If they have a significant cosmological density,
couple to "acoustic" gravitation and have accreted with "acoustic" matter,
they should be present in terrestrial materials. However, it should be noticed
that the gravitational force they would feel on the surface of earth is
$\approx 10^{12}~eV~m^{-1}~G_{O,a}~G_N^{-1}$ for a mass 
$M_O~\approx 10^{-5}~gm$ (the mass scale suggested by the above
considerations), $G_{O,a}$ being the gravitational coupling constant between
"acoustic" and "optical" particles, and $G_N$ Newton's constant.
If $G_{O,a}~G_N^{-1}$ is not much below 1 , this force may have been strong 
enough to favour the diffusion of "optical" particles trough terrestrial
matter towards the center of the earth.
Dedicated experiments could, for instance, search in fine powders for 
very heavy particles
with unconventional responses to external forces.  
\vskip 5mm
{\bf 1c. The low-energy limit}
\vskip 5mm
The above analogies remind simple, well-known examples of how dynamics at
a small scale of distance (the interatomic distance) with  
energy scale 
$\approx ~1~keV$ ($T~\approx ~10^7~K$) can generate, 
and naturally protect,
objects (the phonons) 
existing as massless excitations 
down to microkelvin temperatures ($\approx 10^{-10}~eV$).
A new dynamical interpretation of massless particles and gauge 
interactions may potentially emerge from such analogies, with inner vacuum
dynamics replacing the standard geometric formulation of the gauge principle
and chiral symmetries.
For instance, rather than associating gauge fields to a 
mathematical way to compare
local definitions of space-time or internal symmetry frames, we can admit 
that there is a well-defined physical 
way to make such comparisons but that the local
frames and parameters
fluctuate. Because of nearest-neighbour interactions, fluctuating 
dynamical links between sites are generated giving rise to the gauge bosons. 
In the above simple-minded model of links, the masslessness of the gauge
boson would be equivalent to assuming that, whatever its value, no energy is
spent in setting the matrix $u~(n~,~n-1)$ to be different from the identity
provided its value is the same for all links. Gauge fields could
possibly be interpreted as polarization states of vacuum with nearest-neighbour
interaction between polarizations (similar to spin systems).
The concept of "vacuum polarization" already appeared long ago in quantum
electrodynamics without any explicit use of a precise  
vacuum structure [19~,~21] .                   
With more than one space dimension, since the local polarization variable
would naturally be a vector, we can associate it with 
links between sites  but also with the sites 
themselves. In this last case, the above  $u~(n~,~n-1)$ would rather become a 
space-vector $U(N)$ 
matrix ${\vec {\bf P}}~(n)$ describing the polarization state
of each site on the lattice. The use of links or of site variables would 
depend on the details of the dynamics and, possibly, on the
scale at which the lattice is built. More fundamental is the absence of
local elastic constants, such as $\omega _{rest}$ in (1) , in order to
preserve the masslessness of the gauge bosons. 
\vskip 4mm
Apart from gravitational lenses and similar experiments able to feel
the effect of dark matter, 
all the above  
phenomena would remain invisible to low-energy experiments. For instance,
in the ordinary sector with 
$h~k~\approx ~1~keV/c$ and $a~\approx ~10^{-33}~cm$ ,
one has $(k~a)^2~\approx 10^{-52}$ , which is clearly 
much too small to produce any
detectable effect. High-energy experiments would be the only way to possibly 
reach sensitivity to this kind of Lorentz symmetry violation. Thus, 
the impressive bounds derived  
from low-energy experimental tests of Lorentz invariance (e.g. [24 , 25]) are
not incompatible with the proposed scenario.
A basic question can be raised: can we still justify the existence of gravity 
as a gauge interaction, if Lorentz invariance is just a low-energy limit?
In the previous analogy, an answer could be that local fluctuations of the
vacuum rest frame and of local dynamical parameters
exist in any case.
For instance, if $g_{\mu \nu }$ is the effective metric tensor in the low-energy
Lorentz-invariant limit of the equations of motion, a fluctuation of the 
coefficients of the time derivative and of the nearest-neighbour interaction
in (1) amounts to a fluctuation of the metric. More generally, we can consider
(1) as a particular configuration of the equation:
\vskip 3mm
\centerline {$A~d^2/dt^2~~[\phi ~(n)]~+~H~d/dt~[~\phi ~(n+1)~-~\phi ~(n-1)]~-$}
\equation
-~K_{fl}~[2~\phi ~(n)~-~\phi ~(n-1)~-~\phi ~(n+1)]~-~
\omega _{rest}^2~\phi ~~=~~0
\endequation
\par
In the continuum limit, the coefficients $A~=~g_{00}$ , $H~=~g_{01}~
=~g_{10}$ and $-K_{fl}~=~g_{11}$ can be regarded as the matrix elements of
a space-time bilinear metric with equilibrium values: $A~=~1$ , $H~=~0$ and 
$K_{fl}~=~K$ . Then, a small local fluctuation:
\equation
A~=~1~+~\gamma
\endequation
\equation
K_{fl}~=~K~(1~-~\gamma )
\endequation
with $\gamma ~\ll ~1$ would be equivalent to a small, static
gravitational field 
created by a far away source. The "graviton" would remain massless if the
local fluctuations of $A$ , $K_{fl}$ and $H$ ,
$\delta A~(n)$ , $\delta K_{fl}~(n)$
and $\delta H~(n)$ , satisfy equations like (1) without
the rest-frequency term (as is the case for the vibrations of a crystal, where
only nearest-neighbour interactions play a role). 
Then, gravitation would still exist even if Lorentz symmetry 
is no longer an exact linear symmetry. 
The graviton would lead to a long-range force,
as long as global fluctuations of $A$ , $K_{fl}$ and $H$ 
would cost no vibrational energy. No obvious incompatibility seems to arise
between this scenario and the violation of Lorentz symmetry at very high
energy, which is related to the finite value of $a$ and, as we shall see in
the next chapter, can produce detectable effects even at very small values
of $k~a$ . 
\vskip 6mm 
{\bf 2. LORENTZ SYMMETRY VIOLATION IN HIGH-ENERGY PHYSICS}
\vskip 5mm
The study of high-energy phenomena must incorporate the fact that, if Lorentz
invariance is broken, relativistics kinematics can no longer be applied.
As an illustrative example, we shall use the kinematics derived from the
analogy with the one-dimensional monoatomic Bravais lattice. But, obviously,
our basic arguments have a more general validity. 
Several consequences can follow from this modification of the usual
framework:
\vskip 5mm
{\bf 2a. The GZK cutoff does no longer apply}
\vskip 5mm
Assume that, in the vacuum rest frame,
the kinematics of "ordinary" particles  
is indeed given by (10)
with universal values of $K$ and $a$ , and $K^{1/2}~a~=~c$ . 
As an example, we
take $a~\simeq ~10^{-33}~cm$ and $K~\simeq 10^{87}~s^{-2}$ . At the distance
scales associated to the highest cosmic ray energies, i.e. energy
$E~\approx ~10^{20}~eV$ and $(k~a)^2~\approx ~10^{-17}$ , we can expand
$E$ as follows:
\equation
E~~\simeq ~~(2\pi)^{-1}~h~
[K~(k^2a^2~-~k^4~a^4/12)~+~(2\pi)^2~h^{-2}~E_{rest}^2]^{1/2}
\endequation  
and, at high energy, we can write:
\equation
E~~\simeq ~~(2\pi)^{-1}~h~c~k~
[1~+~2\pi ^2~(h~c~k)^{-2}~E^2_{rest}~-~k^2~a^2/24]
\endequation 
\par 
Corrections from Lorentz symmetry breaking are thus of order $10^{-18}$ , i.e. 
$\approx 100~eV$ . This is to be compared with the term from rest energy  
in (36) which, for a proton and at the same energy, is $\approx 10^{-2}~eV$ .
Thus, Lorentz symmetry violation can play an important role in kinematics at
these energies. A proton with $E~>~10^{20}~eV$ interacting with a cosmic
microwave background photon would be sensitive to these corrections.
For instance,
after having absorbed a $10^{-3} eV$ photon moving in the opposite direction,
the proton gets an extra $10^{-3} eV$ energy, whereas its momentum is
lowered by $10^{-3} eV/c$ . In the conventional scenario with exact Lorentz
invariance, this is enough to allow the excited proton to decay into a
proton or a neutron 
plus a pion, losing an important part of its energy. However, it
can be checked that in our scenario with Lorentz invariance violation such
a reaction is strictly forbidden. Writing for the maximum
allowed energy transfer the equation:
\equation
2\pi~(h~c)^{-1}~[E~(k~+~2\pi .10^{-3}~eV/hc)~+~10^{-3}~eV]~-~k~~=~~\delta _p~+
~\delta _{\pi }
\endequation
\noindent
with
\equation
\delta _p~~=~~
(k~-~k_{\pi })~[2\pi ^2~h~^{-2}(k~-~k_{\pi })^{-2}~m_p^2c^2~-
~(k~-~k_{\pi })^2~a^2/24]
\endequation
\equation
\delta _{\pi}~~=~~
~k_{\pi }~
[2\pi ^2~(h~k_{\pi })^{-2}~m_{\pi }^2~c^2~-~k_{\pi }^2~a^2/24]
\endequation
where $m_p$ and $m_{\pi }$ are respectively the proton and the pion mass,
$k_{\pi }$ the momentum of the produced pion,
and the left-hand-side of the equation can be approximated by:
\equation
2\pi~(h~c)^{-1}~[E~(k~+~2\pi .10^{-3}~eV/hc)~+~10^{-3}~eV]~~
\approx ~~E~(k)~+~2.10^{-3}~eV
\endequation
it turns out that corrections due to Lorentz symmetry violation completely
preclude
the reaction. Elastic $p~+~\gamma $ scattering is permitted, but allows the
proton to release only a small amount of its energy, as can be seen
writing:
\equation
2\pi~(h~c)^{-1}~[E~(k~+~2\pi .10^{-3}~eV/hc)~+~10^{-3}~eV]~-~k~~=~~\delta _p~+
~\delta _{\gamma }
\endequation
\noindent
with $\delta _p$ defined as before replacing $k_{\pi }$ by the 
outgoing photon wave vector $k_{\gamma }$ and:
\equation
\delta _{\gamma }~~=~~
-~k_{\gamma }^3~a^2/24
\endequation
whose solution is in the range $k_{\gamma }~\approx 10^{-5}~k$ . 
Thus, the outgoing photon energy for an incoming $10^{20}~eV$ 
proton cannot exceede
$\Delta E^{max} ~ \approx 10^{-5}~E~=~10^{15}~eV$ instead of the value
$\Delta E^{max} ~ \approx ~10^{19}~eV$ obtained with exact Lorentz invariance. 
Similar or more stringent bounds exist 
for channels involving lepton production. Furthermore, obvious phase
space limitations will also lower the event rate, as compared to standard
calculations using exact Lorentz invariance which predict photoproduction
of real pions at such cosmic proton energies. The effect seems strong enough
to invalidate the Greisen-Zatsepin-Kuzmin (GZK) cutoff [26] and explain the 
existence of the highest-energy cosmic rays [27] . It will become more
important at higher energies, as we get closer to the $a^{-1}$ wavelength
scale. Similar arguments apply to heavy nuclei, again invalidating the GZK
cutoff. Since, in both cases, the cosmic ray energy was expected to degrade
over distances $\approx 10^{24}~m$ 
according to conventional estimates, the correction by several orders
of magnitude we just introduced applies to distance scales much larger than the
estimated size of the presently observable 
Universe. It is not possible to extend the
argument to photons of $E~\approx 10^{14}~eV$ , because in this case one would
have  $(k~a)^2~\approx ~10^{-29}$ leading to too small corrections.
\vskip 4mm
Thus, compared to the model considered by Coleman and Glashow [15] , the
present scenario (where $K$ and $a$ have an exactly universal value for 
all "ordinary" particles)
produces the reverse effect. Not only the existence of very
high-energy cosmic rays is not an evidence against Lorentz symmetry violation,
but the experimental failure of the GZK cutoff for protons and nuclei
would be an evidence for a deviation from relativistic kinematics.
Obvioulsy, a better understanding of the dynamics at Planck scale is needed.
\vskip 5mm
{\bf 2b. Unstable high-momentum particles live longer than naively expected}
\vskip 5mm
In standard relativity, we can compute the lifetime of any unstable particle
in its rest frame and, with the help of a Lorentz transformation, obtain
the Lorentz-contracted lifetime for a particle moving at finite speed. This
is no longer possible with the kinematics defined by (10). If a
particle of mass $m$ and momentum $k$
decays into two particles (particles 1 and 2) with masses $m_1$ and $m_2$ ,
$m~>~m_1~+~m_2$ , we can write at high energy,
in the energetically most favourable configuration (no transverse energy):
\equation
E~(k~,~m)~~=~~E~(k'~,~m_1)~+~E~(k~-~k'~,~m_2)
\endequation
\noindent
which, with a simple change of variables using (10) can be turned into:
\equation
[sin^2~(\theta /2)~+~\alpha ]^{1/2}~~=~~
[sin^2~(\theta _1/2)~+~\alpha _1]^{1/2}~+~
[sin^2~(\theta _2/2)~+~\alpha _2]^{1/2}
\endequation
\noindent
where $\theta ~=~k~a$ , $\alpha ~=~(\pi ~m~c~a)^2~h^{-2}$ and similarly
for $\theta _1$ , $\theta _2$ , $\alpha _1$ and $\alpha _2$ . Writing for
simplicity $\alpha _1~=~\alpha _2$ and setting 
the configuration
$\theta _1~=~\theta _2$ , (always allowed with exact Lorentz invariance
in which case it requires nonzero transverse energy),
the equation becomes:
\equation
sin^2~(\theta /2)~+~\alpha~~=~~4~[sin^2~(\theta _1/2)~+~\alpha _1]
\endequation
\noindent
and, for $\theta _1~=~\theta _2/2$ :
\equation
1~-~cos~(k~a/2)~~=~~(\alpha ~-~4~\alpha _1)^{1/2}
\endequation
\noindent
and, for $\alpha $ , $\alpha _1$ and $k~a$ $\ll ~1$ , can be approximated by:
\equation
k~~\simeq ~~2^{1/2}~(\pi ~c)^{1/2}~(a~h)^{-1/2}~(m^2~-~4~m_1^2)^{1/4}
\endequation
so that the configuration is forbidden above this value of $k$ . This implies 
an important reduction of phase space for the decay of very
high-energy particles:
their lifetimes get longer than it was expected with exact Lorentz invariance.
More complicate expressions with similar 
meaning can be derived for $m_1~\neq ~m_2$ . If $m_1$ and $m_2$ are both 
nonzero, a stronger constraint can be obtained which forbids the decay at very
high energy. To derive it, we can write: $\theta ~=~\theta _1~+~
\theta _2$ , and expand equation (44). We are thus led to:
\equation
\alpha ~-~\alpha _1 ~-~\alpha _2~~=~~2~[sin^2~(\theta _1/2)~
sin^2~(\theta _2/2)~+~D~(\theta _1 , \theta _2 , \alpha _1 , \alpha _2)]
\endequation
where: 
\vskip 3mm
\centerline {$D~(\theta _1 , \theta _2 , \alpha _1 , \alpha _2)
~~=~~[sin^2~(\theta _1/2)~
+~\alpha _1]^{1/2}~[sin^2~(\theta _2/2)~+~\alpha _2]^{1/2}~-$}
\equation
~-~
sin~(\theta _1/2)~sin~(\theta _2/2)~cos~(\theta _1/2)~cos~(\theta _2/2)
\endequation
\par
If $m_1$ or $m_2$ vanishes, it will always be possible to keep the left-hand
side of (48) larger than the right-hand side taking, for instance, $m_1~=~0$
and $sin~(\theta _1/2)$ small enough (although most of the usual phase space 
would then be lost). However, this possibility does no 
longer exist if none of the two masses vanish. In general, any decay with at
least two massive particles being part of 
the final state is forbidden at very high 
energy. We can check the existence of such bounds
minimizing the right-hand side in (48) . 
If $m_2$ is the smallest mass, a typical bound will
forbid the decay for $E~>~E^{st}$ where:
\equation
E^{st}~~\approx ~~c^{3/2}~
h^{1/2}~(a~m_2)^{-1/2}~(m^2~-~m_1^2~-~m_2^2)^{1/2} 
\endequation
Thus, 
as a result of Lorentz symmetry violation, unstable particles 
and nuclei may become stable when accelerated to very high momenta, provided
all decay channels contain at least two massive particles. The energy scale
above which the decay is forbidden varies like the inverse square root
of the mass of the lightest particle produced by the decay.
\vskip 4mm
{\it The neutron 
would become stable} for $E~\stackrel{>}{\sim }~10^{20}~eV$ .
At the same energies or slightly above, {\it some unstable
nuclei would also become stable}. 
Similarly, {\it some hadronic resonances} (e.g. the $\Delta ^{++}$ , whose decay
product must contain at least a proton and a positron)
{\it would become stable} at 
$E~\stackrel{>}{\sim }~10^{21}~eV$ .  
Most of these objects will 
decay before they can be accelerated to such energies, but 
they may result of a collision at very high energy or of
the decay of a 
superluminal particle. The study of very high-energy cosmic rays can thus
reveal as stable particles objects which would be unstable if produced at
accelerators. 
\vskip 4mm
Neutrino masses and 
oscillations are also important in order to discuss the lifetimes of
high-energy particles, as we are often confronted to decay modes involving
muon and electron neutrinos. For instance, if one of the light neutrinos
($\nu _e$ , $\nu _{\mu }$) has a mass in the $\approx ~10~eV$ 
range, {\it the muon would
become stable} at energies above $\approx 10^{22}~eV$ .
Let $E_1$ , $E_2$ , $m_1$ and $m_2$ be the energies and masses of neutrinos
$\nu _1$ and $\nu _2$ , which mix to give neutrinos $\nu $ and $\nu '$ 
with energies $E$ , $E'$ and masses $m$ , $m'$ .
We write (10) as: $E~=~F~(k~,~m)$ for neutrino $\nu $ , and similar expressions
for the other neutrinos. A simple mixing scheme would be to add 
to the hamiltonian non-diagonal elements:
\equation
<\nu _1\mid H \mid \nu _2>~~=~~<\nu _2\mid H \mid \nu _1>~~=~~\Delta ~(k)
\endequation
and require that the hamiltonian has eigenvalues $E$ and $E'$ given by 
the above described expressions. We then get:
\equation
\Delta ^2~(k)~~=~~F~(k~,~m_1)~F~(k~,~m_2)~-~F~(k~,~m)~F~(k~,~m')
\endequation
\noindent
and a mixing angle $\psi $ given by:
\equation
tan ~(2~\psi )~~=~~2~\Delta (k)~[F~(k~,~m_1)~-~F~(k~,~m_2)]^{-1}
\endequation
\par
Going back to a more fundamental level, if $\phi _1$ and $\phi _2$ 
are two different
(but related) degrees of freedom satisfying equation (1) 
and describing $\nu _1$ and $\nu _2$ with
the same value of $K$ and different values of $\omega ^2_{rest}$ , adding
a term proportional to 
$\mid \phi _1~-~\phi_2 \mid ^2$ to the lagrangian would indeed
result
in a pure mass mixing of the type we just presented
{\it with constant $\psi $} , without changing the
effective
value of $K$ (therefore leaving the critical speed in vacuum unchanged).
However, since Lorentz invariance does no longer 
hold, the form of the dispersion
relation (10) may be modified at high wavelength by anharmonic effects.
Mixing with superluminal particles would produce 
different effects, as it necessarily
implies particles associated to oscillations with different values of $K$ and
cannot leave sectorial kinematics unchanged.  Such
a mixing can strongly modify the parameters of our above discussion 
(see also [11] for an explicit example), 
but in any case it contributes to invalidate the GZK cutoff
and to possibly permit unstable particles to become stable at high momentum. 
\vskip 6mm
{\bf 3. SUPERLUMINAL PARTICLES AND STANDARD COSMOLOGY}
\vskip 5mm
It is well known that,
in the standard Big Bang model without inflation [17~,~18] , the horizon 
problem arises basically from the fact that the most distant sources we
can observe now with microwave antennae pointing in opposite directions
must have been $\approx 100$ horizon lengths apart when the cosmic 
microwave background radiation decoupled ($T$ , temperature, 
$\approx 3~.~10^3~K~$ ,
$t$ , age of the Universe, $~\approx ~10^{13}~s$). Given the observed 
isotropy of cosmic background radiation, it seems difficult to understand how 
regions of the Universe that were not in causal contact could have acquired
the same temperature up to $\approx ~10^{-5}$ fluctuations. 
Temperature fluctuations are related to density fluctuations which were 
in principle generated much earlier in the history of the Universe.
\vskip 4mm
Superluminal particles may provide a natural alternative to inflation in order 
to solve the horizon problem.
For instance, assuming that the size of the 
presently observable Universe is $\approx 10^{26}~m$ 
and its age $\approx 10^{17}s$ , 
and a standard evolution for
$R$ (the cosmological distance scale measured by the radius of
the presently observable Universe), we find a ratio $R~t^{-1}~\approx ~10^4~c$
at cosmic time 
$t~\approx ~10^7~s$ and $k_B~T~\approx 1~keV$ . A superluminal particle
with critical speed $c_i~\approx ~10^{12}~c$ 
and rest energy $E_{rest}~=~mc_i^2~
\approx 100~GeV$ ($m~\approx 10^{-13}~eV~c^{-2}$)
in thermal equilibrium (i.e. with $v~\approx ~10^8~c$) 
can cross the Universe provided it can keep most of
its energy for more than $\approx ~10^7~s$ . 
The effect of collisions has been 
taken into account, assuming that the particle
undergoes a few scatterings per second.
This scenario requires "Cherenkov" 
radiation in vacuum to be very weak. At such energies, the emitted 
ordinary particles
woud necessarily be photons and neutrino-antineutrino pairs. Therefore, the
superluminal particle should have weak enough electroweak couplings. 
By traveling at very high speed, such particles can 
emit pairs of "back-to-back" ordinary photons 
and neutrinos (see, e.g. [9~,~10]) or locally
thermalize slower (therefore, heavier) superluminal particles which in turn 
would thermalize ordinary matter. In this way, it may be possible to solve the 
horizon problem and simultaneously find dark matter candidates, if superluminal
particles are abundant enough to compensate the expected weak  
coupling between different sectors. More precise considerations would require
building a global cosmological model, of which we give some possible 
ingredients below. 
\vskip 4mm
To attempt a description of the cosmological role 
of superluminal particles,
we assume that a theory of all gravitation-like forces can be
built, taking at each point the 
vacuum rest frame, and generalize Friedmann equations writing for
a flat Universe in the present epoch (where 
pressure can be neglected):
\equation
R^{-1}~d^2R/dt^2~~\approx ~~- 4\pi ~ Z_2~Z_1^{-1}/3~+~\Lambda /3
\endequation
\equation
(R^{-1}~dR/dt)^2~~\approx ~~ 8\pi ~Z_2~Z_1^{-1}/3~+~\Lambda /3
\endequation
where $\Lambda $ is the cosmological constant [17~,~18] and, in a simplified 
scheme:
\equation
Z_1~~=~~\rho _a~+~\rho _O ~+~\Sigma _i~(\rho _{a,i}~+~
\rho _{O,i})~
\endequation
\equation
Z_2~~=~~G_a~\rho _a^2~+~G_O~\rho _O ^2~+~\Sigma _i~(G_{a,i}~\rho _{a,i}^2~+~
G_{O,i}~\rho _{O,i}^2)
\endequation
\noindent
where $G_a~=~G_N$ is Newton's gravitational constant, 
$\rho _a$ the density of "acoustic" ordinary matter,
$\rho _O$ the density of "optical" ordinary matter
(taken to be positive), $\rho _{a,i}$ and
$\rho _{O,i}$ the densities of "acoustic" and "optical" matter of the $i$-th
superluminal sectors (again, taking the densities of "optical" particles
to be positive), and the $G$'s are effective gravitation-like
coupling constants.
$Z_2~Z_1^{-1}$ replaces the usual expression $G_N~\rho $ in standard 
Friedmann equations.
$Z_1$ is the total density of "particle matter",
where the expression "particle matter" 
designs all 
possible excitations of vacuum that we can describe as particles. 
\vskip 4mm
Expressions (54) to (57) can be derived, for instance, by associating
standard Friedmann equations (with only one "gravitational" component)
to a lagrangian in terms of $R$ and $dR/dt$ , and generalizing the
expressions for kinetic and potential energies in the limit where gravitational
couplings between different components of $Z_1$ are small. An interaction
between the different "gravitational" components of $Z_1$ is, 
even in this case, 
implicitly generated by the constraint that $R$ and $dR/dt$ are space-time
variables common to all the kinds of matter we consider. 
Since, at the same time, cosmology considers space-time as being
generated by matter, this is indeed an effective dynamical interaction
between matter from different sectors. The role of
vacuum is crucial in the generation of a single, absolute space-time
with a local absolute rest frame. 
\vskip 4mm
The new formulae reflect the fact that the "graviton" coupled to
superluminal matter is in principle not the same which couples to "ordinary" 
matter (indeed, local fluctuations of the vacuum rest frame 
and dynamical constants can be 
sector-dependent), assume for simplicity a similar separation between 
"optical" and "acoustic" matter inside the same sector
and, to a first approximation, neglect effects due to the 
mixing between the effective gravitons coupled to different components.
Modifications to this schematic description can be readily introduced, but
would not change our basic reasoning and conclusions concerning the new
flexibility of cosmological fits and the allowed values of the cosmological 
constant. For instance, we can add to $Z_2$ "non-diagonal" terms due to
gravitation-like interactions between different components of $Z_1$ .
Defining, as usual [17~,~18], the critical density $\rho _c$ as
the "acoustic" ordinary matter density which, alone, would make the 
standard Friedmann equations compatible with a flat Universe 
without cosmological constant and with the
measured value of Hubble's constant, i.e. $\rho _c~=~3~(8\pi~G_N)^{-1}~H_0^2$
where $H_0$ is the current value of $R^{-1}~dR/dt$ , we can write:
\equation
\Omega _{\Lambda}~~=~~(8\pi~G_N~\rho _c)^{-1}~\Lambda
~~=~~(3~H_0^2)^{-1}~\Lambda
\endequation
\equation
\Omega _{\alpha }~=~
G_{\alpha }~\rho  _{\alpha }^2~(G_N~\rho _c~Z_1)^{-1}~~=~~
(3~Z_1~H_0^2)^{-1}~G_{\alpha }~\rho _{\alpha }^2 
\endequation
where $G_{\alpha }~\rho  _{\alpha }^2$ is one of the components of $Z_2$ in
(57) , i.e. $\alpha ~=~a~,~O~,~(a~,~i)~,~(O~,~i)$ . We then get, for a flat
Universe, the relation:
\equation
\Omega ~=~\Omega _{\Lambda }~+~\Sigma _{\alpha }~\Omega _{\alpha }~=~1
\endequation
\par
In the recent years, there have been claims [28~,~29] 
in favour of a comparatively large
cosmological constant which could amount, using the above 
definitions of the $\Omega $-like parameters, to values of 
$\Omega _{\Lambda }$ (the contribution of the cosmological constant to the  
expansion of the Universe) as large as $\approx ~0.6$.
The scenario we propose, with many components of $Z_1$ and $Z_2$ , would allow
for rather small values of $Z_2~Z_1^{-1}$ (typically, if there
are many components with similar weights
and weakly interacting with eachother), and therefore possibly
for $\Omega _{\Lambda }$ naturally close to 1 
in a flat Universe, even if $\rho _a/\rho _c$ 
is equal to
0.3 or larger. 
Thus, at the price of considerably weakening the connection
between the density of
ordinary "acoustic" matter and the parameters governing the expansion
of a flat Universe but possibly getting closer to reality,
the new formulation would really make easier 
cosmological fits willing to simultaneously describe matter in
the Universe, galaxy formation, the
age of the Universe, and the spectrum 
and isotropy of cosmic microwave background.
If the Universe is not flat, the following term accounting for cosmic curvature
should 
be added to the right-hand side of (55):
\equation
Z_{curv}~~
=~~-~k_c~R_U^{-2}~Z_1^{-1}~\Sigma _{\alpha }~\rho _{\alpha }~c_{\alpha }^2
\endequation
where $k_c~=~\pm~1$ is the curvature constant, $R_U$ the curvature radius
of the Universe (most likely, $R_U~\gg ~R)$
and $c_{\alpha }$ the critical speed of 
each component of $Z_1$ . From (54~-~57), the formula for the deceleration 
parameter $q$ [17~,~18] becomes:
\equation
q~~=~~-~(R~d^2R/dt^2)~(dR/dt)^{-2}~~=~~H^{-2}~
(4\pi ~Z_2~Z_1^{-1}~-~\Lambda )/3
\endequation
where $H~=~R^{-1}~dR/dt$ is Hubble's "constant" and, writing:
\equation
-~Z_{curv}~~=~~-~H^2~+~(8\pi ~Z_2~Z_1^{-1}~+~\Lambda )/3~=
~H^2~(2~q~-~1)~+~\Lambda
\endequation
we generalize the well-known relation between curvature, deceleration and
Hubble's constant [17~,~18] in the presence of a nonvanishing 
cosmological constant. Defining $\Omega $ as before, expression
(60) applied to the present Universe becomes now:
\equation
\Omega ~~=~~\Omega _{\Lambda }~+~\Sigma _{\alpha }~\Omega _{\alpha }~~=
~~1~-~Z_{curv }~H_0^{-2}~~=
~~1~+
~k_c~R_U^{-2}~H_0^{-2}~Z_1^{-1}~\Sigma _{\alpha }~\rho _{\alpha }~c_{\alpha }^2~
\endequation
or, eliminating $Z_{curv}$ in terms of the present value of $q$ , $q_0$ :
\equation
\Omega _{particles }~~=~~
\Sigma _{\alpha }~\Omega _{\alpha }~~=~~2~(q_0~+~\Omega _{\Lambda})
\endequation
which is similar to standard formulae, but with the definitions
(58) and (59). Whether 
the Universe is flat or curved, the equalities (64) and (65) lead to
the standard relations:
\equation
\Omega _{\Lambda }~~=~~(1~-~2q_0~-~Z_{curv }~H_0^{-2})/3
\endequation
\equation
\Omega _{particles}~~=~~2~(1~+~q_0~-~Z_{curv }~H_0^{-2})/3
\endequation
\noindent
with $Z_{curv }~=~0$ in a flat Universe. The requirement that $\Omega
_{particles}$ be positive, combined with experimental bounds on $q_0$ , 
puts bounds on a positive value of $Z_{curv }$ (corresponding to $k~=~-1$).
The situation seems less obvious for negative values of $Z_{curv }$ ,
if superluminal particles exist. In recent fits [28~,~29] , $\Omega _{\Lambda }$
tends to get close to its maximum value compatible with experimental lower
bounds on $q_0$ , and a negative $Z_{curv }$ would allow for a larger 
$\Omega _{\Lambda }$ provided the contribution of superluminal sectors 
to $\Omega _{particles }$ is large enough.
In (54) and (55), the cosmological constant $\Lambda $ is actually:
\equation
\Lambda~~=~~Z_1^{-1}~\Sigma _{\alpha }~\Lambda _{\alpha }~\rho _{\alpha }
\endequation
where the $\Lambda _{\alpha }$ are sectorial cosmological constants, varying
much slower than $R$ . The expansion of the Universe seems thus
to generate the vacuum matter which produces the sectorial cosmological
constants.
It is quite naturally that a significant cosmological constant would 
arise in our 
approach. If $Z_1$ and $Z_2$ describe the role of vacuum
excitations, but the expansion of the Universe 
is still generating the matter which forms
the ground state of vacuum (i.e. "vacuum" itself), 
this evolution is expected to spend 
a sizeable amount of energy in the creation of new matter:
it must be driven by vacuum dynamics and vacuum energy. Then, the
presence in (54) and (55) of terms describing inner vacuum dynamics seems
compelling, even at a deeper level than for inflationary
models. 
For instance, vacuum may have two sets of degrees of freedom: a) one, 
presently at low temperature, which produces all the objects that we call
"particles"; b) a second one at higher temperature (possibly undergoing  
a second-order phase transition and weakly coupled to the 
particles we observe), whose cooling generates 
vacuum matter and drives vacuum expansion. It is well kown, in
condensed matter physics, that two weakly interacting sets of degrees of
freedom can remain for a long time at different temperatures
(e.g. in adiabatic demagnetization).
It is not obvious how well the simplified approach we
adopted allows to describe the possibly complex role of vacuum, 
as a dynamical system with many degrees of freedom, in
the present expansion of the Universe. Most likely, 
in spite of the important successes of present models
[30~, 31] , crucial ingredients 
describing the role of inner vacuum dynamics are still missing in standard 
cosmology. Current "Pre-Big Bang" cosmology [32] is based on a 
superstring approach where
vacuum dynamics is accounted for 
by the spectrum and properties of the complete set of
vacuum excitations described by the superstrings. 
Alternatives approaches could be directly based on
inner vacuum dynamics, naturally generating superluminal particles and
Lorentz symmetry violation. 
\vskip 1cm
{\bf Acknowledgements}
\vskip 5mm 
It is a pleasure to thank Dr. J. Gabarro-Arpa, as well as colleagues at LPC
(Coll\`ege de France), for useful discussions. 
\vskip 1cm
{\bf References}
\vskip 5mm
\noindent
[1] S.K. Lamoreaux, Phys. Rev. Lett. 78 , 5 (1997); M.J. Sparnaay, Physica 24 ,
751 (1958).
\par
\noindent
[2] H.B.G. Casimir, Koninkl. Ned. Akad. Wetenschap. Proc. 51 , 793 (1948).
\par
\noindent
[3]
E. Elizalde and A. Romeo, Am. J. Phys. 59 , 711 (1991).
\par
\noindent
[4] V.M. Mostepanenko and N.N. Trunov, Sov. Phys. Usp. 31 , 965 (1988).
\par
\noindent
[5] L. Gonzalez-Mestres, "Properties of a possible class
of particles able to travel faster
than light", Proceedings of the Moriond Workshop on "Dark Matter in Cosmology,
Clocks and Tests of Fundamental Laws", Villars (Switzerland), January 21-28
1995 , Ed. Fronti\`eres, France. Paper astro-ph/9505117 of electronic library.
\par
\noindent
[6] L. Gonzalez-Mestres, "Cosmological implications of a possible class of
particles able to travel faster than light", Proceedings of the Fourth
International Workshop on Theoretical and Experimental Aspects of
Underground Physics, Toledo (Spain) 17-21 September 1995~, Nuclear Physics B
(Proc. Suppl.) 48 (1996). Paper astro-ph/9601090 .
\par
\noindent
[7] L. Gonzalez-Mestres, "Superluminal matter and high-energy cosmic rays",
May 1996 . Paper astro-ph/9606054 .
\par
\noindent
[8] L. Gonzalez-Mestres, "Physics, cosmology and experimental signatures of
a possible new class of superluminal particles", to be published in the
Proceedings of the International Workshop on the Identification of Dark
Matter, Sheffield (England, United Kingdom), September 1996 . Paper
astro-ph/9610089 .
\par
\noindent
[9] L. Gonzalez-Mestres, "Physical and cosmological implications of a possible
class of particles able to travel faster than light", contribution to the
28$^{th}$ International Conference on High-Energy Physics, Warsaw July 1996 .
Paper hep-ph/9610474 .
\par
\noindent
[10] L. Gonzalez-Mestres, "Space, time and superluminal particles", February
1997 . Paper mp$_-$arc 97-92 and physics/9702026 .
\par
\noindent
[11]  L. Gonzalez-Mestres, "Lorentz invariance and superluminal particles", 
March 
1997 . Paper mp$_-$arc 97-117 and physics/9703020 .
\par
\noindent
[12] See, for instance, N.W. Ashcroft and N.D. Mermin, "Solid State Physics",
Saunders College, Philadelphia, USA 1976 .
\par
\noindent
[13] See, for instance,
"Tachyons, Monopoles and Related Topics", Ed. by E. Recami,
North-Holland 1978 , and references therein.
\par
\noindent
[14] 
The relativity principle was formulated by
H. Poincar\'e, Speech at the St. Louis International Exposition of 1904 ,
The Monist 15 , 1 (1905).
\par
\noindent
[15] 
S. Coleman and S.L. Glashow, "Cosmic ray and neutrino tests of
special relativity", Harvard preprint HUTP-97/A008 , paper
hep-ph/9703240 .
\par
\noindent
[16]
See, for instance, L. Bergstrom et al., "THE AMANDA EXPERIMENT: status and
prospects for indirect Dark Matter detection", same Proceedings as for ref.
[8]. Paper astro-ph/9612122 .
\par
\noindent
[17] See, for instance, P.J.E. Peebles, "Principles of Physical Cosmology",
Princeton Series in Physics, Princeton University Press 1993 .
\par
\noindent
[18] See, for instance, P.D.B. Collins, A.D. Martin and E.J. Squires,
"Particle Physics and Cosmology", Wiley 1989 .
\par
\noindent
[19] S.S. Schweber, "An Introduction to Relativistic Quantum Field Theory",
Row, Peterson and Company 1961 .
\par
\noindent
[20]
R.F. Streater and A.S. Wightman, "$PCT$, Spin and Statistics, and All
That", Benjamin, New York 1964 ; R. Jost, "The General Theory of
Quantized Fields", AMS, Providence 1965 .
\par
\noindent
[21]  C. Itzykson and J.B. Zuber, "Quantum Field Theory", McGraw-Hill 1985 .
\par
\noindent
[22] See, for instance, E. Recami in [7] .
\par
\noindent
[23] See, for instance, E.C.G. Sudarshan in [7] .
\par
\noindent
[24] S.K. Lamoreaux, J.P. Jacobs, B.R. Heckel, F.J. Raab and E.N. Forston,
Phys. Rev. Lett. 57 , 3125 (1986); D. Hils and J.L. Hall, Phys. Rev. Lett.
64 , 1697 (1990).
\par
\noindent
[25] M. Goldhaber and V. Trimble, J. Astrophys. Astr. 17 , 17 (1996).
\par 
\noindent
[26] K. Greisen, Phys. Rev. Lett. 16 , 748 (1966); G.T. Zatsepin and V.A. 
Kuzmin, Pisma Zh. Eksp. Teor. Fiz. 4 , 114 (1966).
\par
\noindent
[27] See, for instance, Proceedings of TAUP 95 , Nuclear Physics B (Proc.
Suppl.) 48 (1996), May 1996 and references therein.
\par
\noindent
[28] See, for instance, M.S. Turner, 
"The Case for $\Lambda $CDM", to be published 
in "Critical Dialogues in Cosmology", Ed. N. Turok, World Scientific Pub.
1997 .
\par 
\noindent
[29] M.S. Turner, "The Cosmology of Nothing", in 
"Vacuum and Vacua: the Physics of Nothing", Ed. A. Zichichi, World
Scientific Pub. 1996 , 
and references therein.
\par
\noindent
[30] M.S. Turner, "Cosmology: Standard and Inflationary", in "Particle Physics, 
Astrophysics and Cosmology", XXII$^{th}$ SLAC Summer Institute, Ed.
J. Chan and L. De Porcel (Conf-9408100 ; National Technical Information
Service, U.S. Department of Commerce, 1994) , and references therein.
\par
\noindent
[31]  M.S. Turner, "Cosmology 1996" , Proceedings of the Fourth KEK Topical
Conference, Tsukuba (Japan) October 1996 . Paper astro-ph/9704024 .
\par
\noindent
[32] G. Veneziano, "Inhomogeneous Pre-Big Bang Cosmology", CERN preprint
TH/97-42 (1997), paper hep-th/9703150 , and references therein.
\end{document}